\documentclass{aastex}
\usepackage{spr-astr-addons}
\usepackage{url}
\usepackage{graphicx}

\RequirePackage{color}



\begin{document}

\title{Radiation Hydrodynamical Instabilities in Cosmological \\
and Galactic Ionization Fronts}
\slugcomment{Not to appear in Nonlearned J., 45.}
\shorttitle{Primordial and Galactic I-fronts}
\shortauthors{Autors et al.}

\author{Daniel J. Whalen\altaffilmark{1}} \and \author{Michael L. Norman\altaffilmark{2}}
\email{dwhalen@lanl.gov}

\altaffiltext{1}{Department of Physics, Carnegie Mellon University, Pittsburgh, PA 15213}
\altaffiltext{2}{Center for Astrophysics and Space Sciences,
University of California at San Diego, La Jolla, CA 92093}

\begin{abstract}
Ionization fronts, the sharp radiation fronts behind which H/He ionizing photons 
from massive stars and galaxies propagate through space, were ubiquitous in the 
universe from its earliest times.  The cosmic dark ages ended with the formation 
of the first primeval stars and galaxies a few hundred Myr after the Big Bang. 
Numerical simulations suggest that stars in this era were very massive, 25 - 500 
solar masses, with H II regions of up to 30,000 light-years in diameter. We 
present three-dimensional radiation hydrodynamical calculations that reveal that 
the I-fronts of the first stars and galaxies were prone to violent instabilities, 
enhancing the escape of UV photons into the early intergalactic medium (IGM) and 
forming clumpy media in which supernovae later exploded.  The enrichment of such 
clumps with metals by the first supernovae may have led to the prompt formation 
of a second generation of low-mass stars, profoundly transforming the nature of 
the first protogalaxies. Cosmological radiation hydrodynamics is unique because 
ionizing photons coupled strongly to both gas flows and primordial chemistry at
early epochs, introducing a hierarchy of disparate characteristic timescales 
whose relative magnitudes can vary greatly throughout a given calculation.  We 
describe the adaptive multistep integration scheme we have developed for the 
self-consistent transport of both cosmological and galactic ionization fronts.
\end{abstract}

\section{First Light and the Rise of Primeval Galaxies}

State of the art numerical simulations reveal that the first stars in the universe 
formed in isolation in small cosmological halos (spheroidal bound clumps of dark 
matter and H/He gas) at redshifts $z \sim$ 20 - 30, or 200 - 400 Myr after the Big 
Bang \citep{abn02,bcl02,nu01}. Soon thereafter, gravitational mergers congregated 
these star forming halos into the first primitive galaxies. Preliminary results from 
minihalo collapse models suggest that primordial (or Pop III) stars were very massive, 
25 - 500 M$_{\odot}$ \citep{on07}.  They created H II regions that were enormous in 
comparison to those in the Galaxy today, 2.5 - 5 kpc in radius \citep{wan04,ket04,abs06,
awb07}.  Primeval ionization fronts, or I-fronts, were key to the rise of the first 
protogalaxies for several reasons. First, Pop III stars lacked strong winds, so their 
H II regions determined their circumstellar environments.  They regulated the flow of 
the first heavy elements from primordial supernovae (SNe) into the early IGM \citep{
get07,get10} and may even have resulted in the prompt formation of a second generation 
of low-mass stars in SNe remnants \citep{wet08b}.  Second, the minihalos forming the 
first stars themselves coalesced in small swarms due to cluster bias. Radiation fronts 
from one star in the cluster could engulf nearby halos, photoevaporating them to varying 
degrees and either suppressing or allowing new star formation in them \citep{sir04,i05,
su06,wet08a,su09,wet10}. Thus, cosmological I-fronts in part governed the number and 
character of stars that were swept up into the first primitive galaxies.

Once primeval galaxies formed and sustained cyclical star formation, they too
propagated I-fronts into the IGM, beginning the process of cosmological reionization 
by which the cold, neutral, mostly featureless universe of the recombination era was 
gradually transformed into the hot, transparent cosmic web of galaxies and clusters
of galaxies we observe today \citep{rgs02,rgs08,wa08,wc09}.  Recent numerical models 
have demonstrated that the I-fronts of both individual Pop III stars and protogalaxies 
formed H$_2$ in their outer layers, a key coolant at high redshifts that may have 
seeded the formation of additional structure in the early universe \citep{rgs01,wn08b}. 
High-redshift I-fronts thus influenced the rise of structure on a variety of spatial 
scales in the early cosmos.  

In the Galaxy today, I-fronts from massive stars in the molecular cores of giant 
molecular clouds (GMCs) are thought to trigger new star formation by perturbing
nearby cores on the verge of gravitational collapse \citep{da07,gr09}.  In this 
manner, successive fronts can set off a chain of star formation throughout the 
cloud.  On the other hand, dynamical instabilities in the fronts may degenerate 
into turbulent flows that support gas against collapse, offsetting triggered star 
formation. Yet, these same instabilities can form clumps that may later collapse 
into stars.  How ionized flows regulate star formation in the Milky Way today 
remains to be fully understood.

\section{Ionizing Radiation Hydrodynamics in ZEUS-MP} 

We perform our ionization front calculations with ZEUS-MP \citep{wn06}, a massively
parallel Eulerian astrophysical hydrodynamics code that solves the equations of ideal 
fluid dynamics\footnote{http://lca.ucsd.edu/portal/codes/zeusmp2}:
\vspace{0.1in}
\begin{eqnarray}
\frac{\partial \rho}{\partial t}  & = & - \nabla \: \cdotp \; (\rho {\bf v})  \\
\frac{\partial \rho v_{i}}{\partial t}  & = & - \nabla \: \cdotp \; (\rho v_{i} 
{\bf v}) \: - \: \nabla p \: - \: \rho \nabla \Phi \: - \: \nabla \cdotp {\bf Q}    \\ 
\frac{\partial e}{\partial t}  & = & - \nabla \: \cdotp \; (e {\bf v}) \: - \: p\nabla \: 
\cdotp \: {\bf v} \: - \: \bf{Q} : \nabla  {\bf v}. \\
&  &  \nonumber
\end{eqnarray}  
Here, $\rho$, $e$, and the $v_{i}$ are the mass density, internal energy density, 
and velocity at each mesh point and $p \;= (\gamma-1) e$ and {\bf{Q}} are the 
gas pressure and the von Neumann-Richtmeyer artificial viscosity tensor. ZEUS-MP 
evolves these equations with a second-order accurate monotonic advection scheme in 
one, two, or three dimensions on Cartesian (XYZ), cylindrical (ZRP), or spherical 
polar (RTP) coordinate meshes.  Our augmented version of the publicly-available 
code self-consistently couples primordial gas chemistry \citep{wn08a} and 
multifrequency photon-conserving UV radiative transfer \citep{wn08b} to fluid 
dynamics for the radiation hydroynamical transport of cosmological I-fronts.

\subsection{Primordial H and He Chemistry}

We evolve H, H$^{+}$, He, He$^{+}$, He$^{2+}$, H$^{-}$, H$^{+}_{2}$, H$_{2}$, and 
e$^{-}$ with nine additional continuity equations and the nonequilibrium rate 
equations of \citet{anet97}: \vspace{0.05in}
\begin{equation}
\frac{\partial \rho_{i}}{\partial t} = - \nabla \: \cdotp \; (\rho_{i} {\bf v}) 
+ \sum_{j}\sum_{k} {\beta}_{jk}(T){\rho}_{j}{\rho}_{k} - \sum_{j} {\kappa}
_{j}{\rho}_{j}, \vspace{0.05in}
\end{equation}
where ${\beta}_{jk}$ is the rate coefficient of the reaction between species j and k 
that creates (+) or destroys (-) species i, and the ${\kappa}_{j}$ are the radiative 
rate coefficients.  We assume that the species share a common velocity distribution. 
Mass and charge conservation, which are not guaranteed by either chemical or advective
updates, are enforced each time the fluid equations are solved.  

Microphysical heating and cooling due to photoionization and gas chemistry is coupled 
to the gas energy density by an isochoric update that is operator-split from updates
to the fluid equations:  
\vspace{0.05in}
\begin{equation}
{\dot{e}}_{\mathrm{gas}} = \Gamma - \Lambda, \label{eqn: egas}
\vspace{0.05in}  
\end{equation}
where $\Gamma$ is the cumulative heating rate due to photons of all frequencies and 
$\Lambda$ is the sum of the cooling rates due to collisional ionization and excitation 
of H and He, recombinations of H and He, inverse Compton scattering (IC) from the CMB, 
bremsstrahlung emission, and H$_2$ cooling.

\subsection{Radiative Transfer}

Our photon-conserving UV transport \citep{mel06,anm99}, which is distinct from the 
flux-limited diffusion native to the public release of ZEUS-MP, solves the static 
approximation to the equation of transfer in flux form to compute radiative rate 
coefficients for the reaction network at every point on the coordinate mesh. As 
currently implemented, our code can transport photons from a point source 
centered in a spherical grid or in plane waves along the $x$ or $z$-axes of Cartesian 
or cylindrical meshes.  To model Pop III spectra, we discretize the blackbody photon 
emission rates of a star with 40 uniform bins from 0.755 to 13.6 eV 
and 80 logarithmically spaced bins from 13.6 eV to 90 eV, normalizing them by the 
total ionizing photon rates by \citet{s02}.  The radiative reactions in our models 
are listed in Table 1 of \citet{wn08b}.  We calculuate H$_2$ photodissociation 
rates along rays parallel to the direction of radiation flow using self-shielding 
functions modified for thermal broadening as prescribed by \citet{db96} to approximate 
the effects of gas motion.  They are shown in equations 9 and 10 of \citet{wn08b}.

\subsection{Radiation Forces}  

Radiation pressure due to ionizations occurs at the I-front itself and in recombining 
gas in the H II region. \citet{wn06} found that the acceleration of fluid elements at 
the front was large but momentary, and only alters the velocity of the front by 1 - 2 
km s$^{-1}$.  Momentum deposition within the H II region is only prominent where gas 
is very dense, like at the center of a minihalo being evaporated by its star.  There, 
rapid successive cycles of ionization and recombination can impart 
radiation forces to the gas that are hundreds of times the strength of gravity at early 
times \citep[lower left panel of Figure 1 in][]{ket04}. As these forces propel gas near 
the center of the halo out into the H II region, its densities and recombination rates 
fall, so more ionizing photons from the star reach the I-front.  This higher flux 
results in I-fronts that are faster than when such forces are not included.  In UV 
breakout from the first star-forming halos, radiation forces speed up the D-type I-front 
by 10 - 20\%. However, this effect is transient: after the internal rearrangement of gas 
deep within the H II region dilutes its interior, radiation forces there sharply fall
 \citep[see also][]{krum09,draine10,lopez10}.  We describe the timescales on which 
 momentum updates are performed below.

\subsection{Adaptive Subcycling}

A hierarchy of highly disparate characteristic time scales arises when gas dynamics, 
radiative transfer, and primordial chemistry are solved in a given application.  The 
three governing times are the Courant time, the chemical time
\begin{equation}
t_{chem} = 0.1 \, \displaystyle\frac{n_{e} + 0.001 n_{H}}{{\dot{n}}_{e}},
\end{equation}
and the photoheating/cooling time \vspace{0.05in}
\begin{equation}
t_{hc} = 0.1 \displaystyle\frac{e_{gas}}{{\dot{e}}_{ht/cool}}.\vspace{0.05in}
\end{equation}
Their relative magnitudes can seamlessly evolve during a single application. 
For example, when an I-front propagates through a medium, photoheating times are 
often smaller than Courant times, and chemical time scales are usually shorter than 
either one. On the other hand, fossil H II regions can cool faster than they 
recombine, so cooling times become shorter than chemical times. The key to solving 
all three processes self-consistently is to evolve each on its own timescale without 
restricting the entire algorithm to the shortest of the times. To successfully deal 
with both I-fronts and relic H II regions, an algorithm must adaptively reshuffle 
the time scales on which the three processes are solved. Implicit schemes are 
sometimes applied to stiff sets of differential equations like those in our 
model because they are unconditionally stable over the Courant time.  However, 
accurate I-front transport in stratified media often requires restricting updates 
to both the gas energy and fluid equations to photoheating times in order to capture 
the correct energy deposition into the gas, and linear system solves over such short 
time scales would be prohibitive in more than one or two dimensions.  Enforcing photon 
conservation in implicit schemes can also be problematic.  

We instead subcycle chemical species and gas energy updates over the minimum of the 
chemical and heating/cooling times until the larger of the two has been crossed, at 
which point we perform full hydrodynamical updates of gas densities, energies, and 
velocities.  These times are global minima for the entire grid.  The chemical times 
are defined in terms of electron flow to accommodate all chemical processes rather 
than just ionizations or recombinations.  Adopting the minimum of the two times for 
chemistry and gas energy updates enforces accuracy in the reaction network when $
t_{chem}$ becomes greater than $t_{hc}$, such as in relic H II regions.  Our 
adaptive subcycling scheme is described in greater detail in \citep{wn08a}.

\begin{figure}[ht]
\begin{center}
\includegraphics[width=19pc]{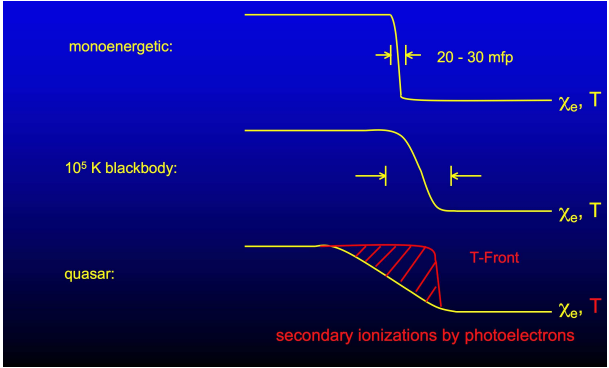}
\end{center}
\caption{\label{fig1}
The structure of an ionization front. $\chi_e$ denotes ionized fraction and T is gas 
temperature.  Top: monoenergetic UV photons just above the ionization edge of H.  
Center: a 10$^5$ K Pop III star blackbody spectrum.  Note the significant  broadening 
of the front.  Bottom: power-law quasar x-ray source spectrum.  Note that in the first two 
cases that temperature tracks the ionization fraction, but not in the quasar case. Here, 
hard photons penetrate well ahead of the front, strongly heating but not significantly 
ionizing the gas.
} 
\end{figure}

\section{Ionization Front Instabilities}

When a point source of monoenergetic ionizing photons appears in a uniform H and 
He medium, they are not free to simply stream into the interstellar medium (ISM) 
because of the large cross section to photoionization in the gas. Instead, they emerge
as a sharp wall of radiation, behind which gas is almost fully ionized and heated
to a few$\times 10^4$ K and beyond which it is completely neutral. The thickness
of the I-front separating the neutral and ionized gas is a few mean free paths of
the ionizing photons through the neutral gas.  The initial propagation of the 
I-front through the gas is supersonic in that the ionization wave advances far 
more quickly than any hydrodynamic response by the hot gas in its wake. This is known
as an R-type I-front.  As the front recedes from the star it slows, partly due to 
geometric dilution of the ionizing flux and partly because recombinations in the 
ionized gas remove photons that would otherwise have gone on to advance the front. 
If the medium is static (no gas motion allowed), the I-front comes to a halt when 
the H II region it bounds encloses enough ionized gas that the global recombination 
rate equals the emission rate of source photons.  The radius at which this occurs 
is known as the Str\"omgren radius.

\begin{figure}[ht]
\begin{center}
\includegraphics[width=19pc]{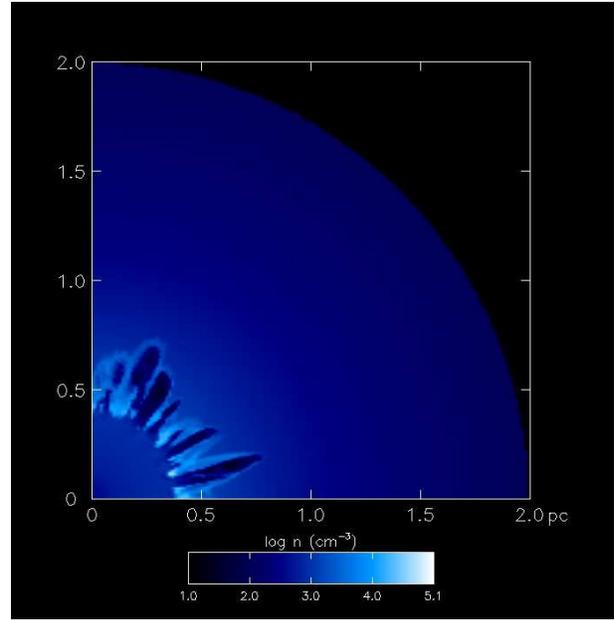}
\end{center}
\caption{\label{fig2}
Thin-shell instability in a primordial I-front mediated by H$_2$ cooling.
} 
\end{figure}

In reality, as the I-front slows to the Str\"omgren radius the pressure wave that
has built up in the hot ionized gas behind it overtakes it and passes through it,
steepening into a shock as it does so.  Thereafter, the H II region expands 
subsonically with respect to the sound speed in its interior, which is still
supersonic with respect to the ambient neutral gas.  As it expands, it bypasses
the static Str\"omgren radius limit because its interior densities and therefore 
recombination rates fall, allowing source photons to continue to stream to the front.  
As the H II region grows it pushes aside more and more ambient gas which accumulates
on its surface, causing the shock to detach from and move ahead of the I-front.  This
is known as a D-type I-front.

\subsection{Thin-Shell Instabilities}

If this shocked neutral shell can radiatively cool, it will collapse into a 
cold dense layer that is prone to fragmentation via Vishniac, or thin-shell, 
instabilities \citep{v83}.  Unlike similar features in the dense shells of 
astrophysical blast waves, these erupt into very violent instabilities because 
ionizing radiation opportunistically escapes through the cracks in the shell 
in jets ahead of the front \citep{gsf96,wn08a}.  These jets rapidly crumple into 
complicated flows that form dense clumps that can persist for the life of the 
star. In the Galaxy today, metals and dust radiatively cool the shell and trigger 
such instabilities.  Cooling by atomic H and He lines in the shocked gas of 
cosmological I-fronts by itself is too weak to incite such instabilities. However, 
they can instead form by H$_2$ cooling.

\begin{figure*}
\epsscale{2.0}
\plotone{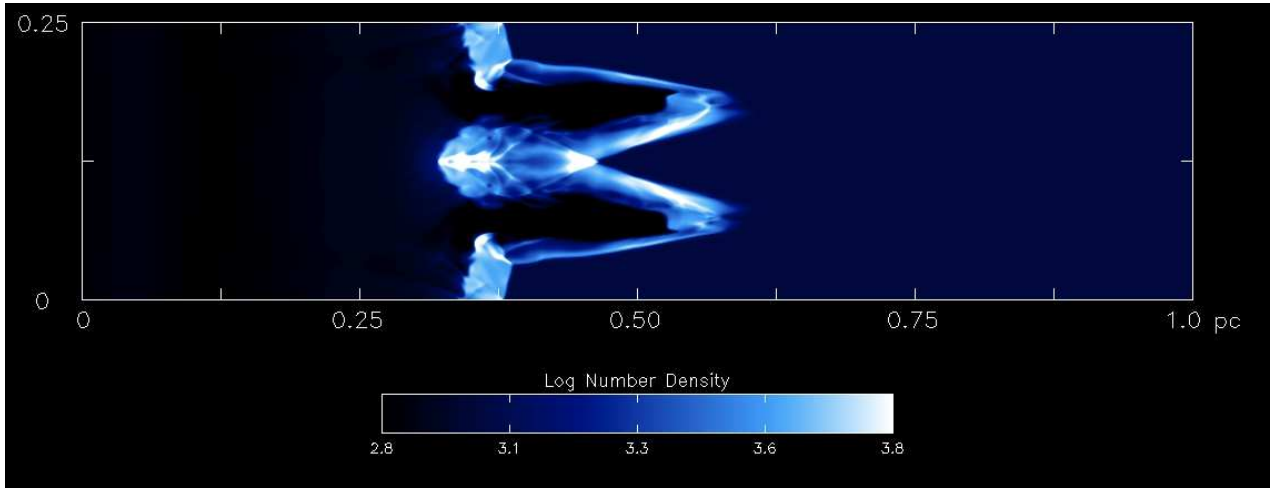}\vspace{0.15in}
\caption{The shadow instability.} 
\vspace{-0.125in}
\end{figure*}

Primordial I-fronts are driven by hard UV sources whose photons have a variety of
mean free paths through neutral gas.  Collectively, they broaden the front, which 
leads to ionized fractions of 10\% and temperatures of 2000 - 3000 K in its outer 
layers as we illustrate in Figure 1.  These are ideal conditions for gas phase 
catalysis of H$_2$ by the H$^-$ and H$_2^+$ channels:
\begin{equation}
H + e^- \rightarrow H^- + \gamma \hspace{0.25in} H^- + H \rightarrow H_2 + e^- 
\end{equation}
\begin{equation}
H + H^+ \rightarrow H_2^+  + \gamma \hspace{0.25in} H_2^+ + H \rightarrow H_2 + H^+ \vspace{0.1in}
\end{equation}
Although much less efficient than metals, H$_2$ is a far more effective coolant
than H and He lines at 3000 K; sandwiched between the front and the shocked neutral
shell, it cools the base of the shell sufficiently to trigger thin-shell
overstabilities that erupt into violent radiation-driven instabilities, as we show
in Figure 2 \citep{wn08b}.  In this simulation, we center a 10$^5$ K blackbody UV
source approximating a 120 M$_{\odot}$ Pop III star in a H/He gas profile similar 
to that of a cosmological minihalo. When the front becomes D-type, instabilities 
due to H$_2$ cooling can be seen to appear at the base of the dense shell.  More 
models are required to determine if the turbulent gas clumps survive until the 
death of the star and are prone to gravitational collapse.

\subsection{Shadow Instabilities}


In Figure 3 we show the breakout of a second kind of radiation hydrodynamical 
instability in primordial I-fronts, the shadow instability \citep{rjw99,wn08b}.
These form when density perturbations are advected through an R-type front. The
perturbation does not trap the front, it merely dimples it.  The walls of the 
dimple are exposed to a smaller ionizing flux because they are oblique to it; as 
a consequence, the walls of the dimple transition to D-type before the tip does,
causing the tip to advance and elongate the dimple into a jet that abruptly breaks 
down into a violent instability.  Such phenomena would have been unavoidable in the 
primeval universe, on scales ranging from UV breakout from the first star-forming
halos to 100 kpc protogalactic I-fronts.  The role of H$_2$ formation, thin-shell
instabilities, and shadow instabilities in structure formation in early cosmological
flows remains to be properly investigated.

\begin{figure}[ht]
\begin{center}
\includegraphics[width=19pc]{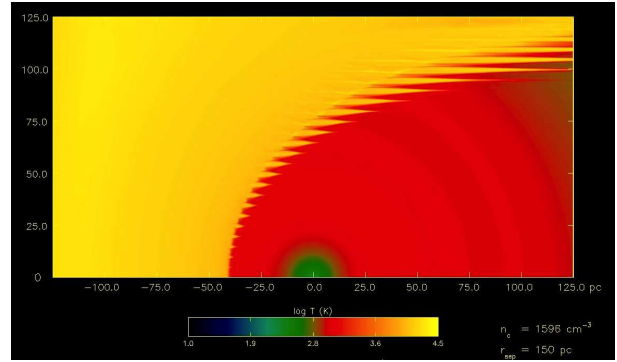}
\end{center}
\caption{\label{fig4}
Angle of incidence (AOI) instabilities in the ionization front of a Pop III star
enveloping a neighbor halo.
} 
\end{figure}

\vspace{-0.05in}

\subsection{AOI Instabilities}

We show in Figure 4 the radiation front of a 25 M$_{\odot}$ Pop III star enveloping
a nearby star-forming halo \citep{wet10}. Spikes in the front can be seen all along 
the arc surrounding the halo, but are increasingly prominent with altitude above its 
axis. This is an example of an I-front instability described by \citet{rjw02}, in 
which photons that arrive at the front at any angle other than 90$^{\circ}$ cause it 
to be unconditionally unstable: an angle of incidence (AOI) instability.  Detailed linear
perturbation analysis predicts that the greater the angle of deviation of the 
photons from pure incidence, the faster the growth of the unstable mode, which true 
of the protrusions in Figure 4.  In this example, the unstable modes have little 
effect on star formation in the halo because those closest to its core have the
smallest amplitudes and fail to puncture it. Dynamical instabilities have also been
found in I-fronts engulfing Galactic molecular cloud cores in numerical models, but 
gas in those simulations could efficiently cool by fine-structure lines \citep{miz05,
miz06} so the unstable modes are likely thin-shell in origin.

\bibliographystyle{spr-mp-nameyear-cnd}

\begin{thebibliography}{}
 
\bibitem[Abel, Norman, \& Madau(1999)]{anm99} Abel, T., Norman, M.~L., \& Madau, P.
\ 1999, \apj, 523, 66 
\bibitem[Abel et al.(2002)]{abn02} Abel, T., Bryan, G.~L., \& Norman, M.~L.\ 2002, 
Science, 295, 93 
\bibitem[Abel et al.(2007)]{awb07} Abel, T., Wise, J.~H., \& Bryan, G.~L.\ 2007, 
\apjl, 659, L87 
\bibitem[Alvarez et al.(2006)]{abs06} Alvarez, M.~A., Bromm, 
V., \& Shapiro, P.~R.
\ 2006, \apj, 639, 621 
\bibitem[Anninos et al.(1997)]{anet97} Anninos, P., Zhang, Y., Abel, T., \& Norman, 
M.~L.\ 1997, New Astronomy, 2, 209 
\bibitem[Bromm et al.(2002)]{bcl02} Bromm, V., Coppi, P.~S., \& Larson, R.~B.\ 
2002, \apj, 564, 23 
\bibitem[Dale et al.(2007)]{da07} Dale, J.~E., Bonnell, I.~A., \& Whitworth, A.~P.
\ 2007, \mnras, 375, 1291 
\bibitem[Draine(2010)]{draine10} Draine, B.~T.\ 2010, arXiv:1003.0474 
\bibitem[Draine \& Bertoldi(1996)]{db96} Draine, B.~T., \&  Bertoldi, F.\ 1996, 
\apj, 468, 269 
\bibitem[Garcia-Segura \& Franco(1996)]{gsf96} Garcia-Segura, G., \& Franco, J.\ 
1996, \apj, 469, 171 
\bibitem[Greif et al.(2007)]{get07} Greif, T.~H., Johnson, J.~L., Bromm, V., \& 
Klessen, R.~S.\ 2007, \apj, 670, 1 
\bibitem[Greif et al.(2010)]{get10} Greif, T.~H., Glover, S.~C.~O., Bromm, V., \& 
Klessen, R.~S.\ 2010, \apj, 716, 510 
\bibitem[Gritschneder et al.(2009)]{gr09} Gritschneder, M., Naab, T., Walch, S., 
Burkert, A., \& Heitsch, F.\ 2009, \apjl, 694, L26 
\bibitem[Iliev et al.(2005)]{i05} Iliev, I.~T., Shapiro, P.~R., \& Raga, A.~C.\ 
2005, \mnras, 361, 405 
\bibitem[Kitayama et al.(2004)]{ket04} Kitayama, T., Yoshida, N., Susa, H., \& 
Umemura, M.\ 2004, \apj, 613, 631 
\bibitem[Krumholz \& Matzner(2009)]{krum09} Krumholz, M.~R., \& Matzner, 
C.~D.\ 2009, \apj, 703, 1352 
\bibitem[Lopez et al.(2010)]{lopez10} Lopez, L.~A., Krumholz, M., Bolatto, A., 
Prochaska, J.~X., \& Ramirez-Ruiz, E.\ 2010, Bulletin of the American Astronomical 
Society, 42, 261 
\bibitem[Mellema et al.(2006)]{mel06} Mellema, G., Iliev, I.~T., Alvarez, M.~A., 
\& Shapiro, P.~R.\ 2006, New Astronomy, 11, 374 
\bibitem[Mizuta et al.(2005)]{miz05} Mizuta, A., Kane, J.~O., Pound, M.~W., Remington, B.~A., 
Ryutov, D.~D., \& Takabe, H.\ 2005, \apj, 621, 803 
\bibitem[Mizuta et al.(2006)]{miz06} Mizuta, A., Kane, J.~O., Pound, M.~W., Remington, B.~A., 
Ryutov, D.~D., \& Takabe, H.\ 2006, \apj, 647, 1151 
\bibitem[Nakamura \& Umemura(2001)]{nu01} Nakamura, F., \& Umemura, M.\ 2001, \apj, 
548, 19
\bibitem[O'Shea \& Norman(2007)]{on07} O'Shea, B.~W., \& Norman, M.~L.\ 2007, \apj, 
654, 66 
\bibitem[Ricotti et al.(2001)]{rgs01} Ricotti, M., Gnedin, N.~Y., \& Shull, J.~M.\ 
2001, \apj, 560, 580 
\bibitem[Ricotti et al.(2002)]{rgs02} Ricotti, M., Gnedin, N.~Y., \& Shull, J.~M.\ 
2002, \apj, 575, 33 
\bibitem[Ricotti et al.(2008)]{rgs08} Ricotti, M., Gnedin, N.~Y., \& Shull, J.~M.\ 
2008, \apj, 685, 21 
\bibitem[Schaerer(2002)]{s02} Schaerer, D.\ 2002, \aap, 382, 28
\bibitem[Shapiro et al.(2004)]{sir04} Shapiro, P.~R., Iliev, I.~T., \& Raga, A.~C.\ 
2004, \mnras, 348, 753 
\bibitem[Susa \& Umemura(2006)]{su06} Susa, H., \& Umemura, M.\ 2006, \apjl, 645, 
L93 
\bibitem[Susa et al.(2009)]{su09} Susa, H., Umemura, M., \& Hasegawa, K.\ 2009, 
\apj, 702, 480 
\bibitem[Vishniac(1983)]{v83} Vishniac, E.~T.\ 1983, \apj, 274, 152 
\bibitem[Whalen, Abel, \& Norman(2004)]{wan04} Whalen, D., Abel, T., \& Norman, 
M.~L.\ 2004, \apj, 610, 14 
\bibitem[Whalen \& Norman(2006)]{wn06} Whalen, D., \& Norman, M.~L. \ 2006, \apjs, 
162, 281 
\bibitem[Whalen et al.(2008a)]{wet08a} Whalen, D., O'Shea, B.~W., Smidt, J., \& 
Norman, M.~L.\ 2008, \apj, 679, 925
\bibitem[Whalen et al.(2008b)]{wet08b} Whalen, D., van Veelen, B., O'Shea, B.~W., 
\& Norman, M.~L.\ 2008, \apj, 682, 49 
\bibitem[Whalen \& Norman(2008a)]{wn08a} Whalen, D.~J., \& Norman, M.~L.\ 2008, 
\apj, 672, 287 
\bibitem[Whalen \& Norman(2008b)]{wn08b} Whalen, D., \& Norman, M.~L.\ 2008, \apj, 
673, 664 
\bibitem[Whalen et al.(2010)]{wet10} Whalen, D., Hueckstaedt, R.~M., \& McConkie, 
T.~O.\ 2010, \apj, 712, 101 
\bibitem[Williams(1999)]{rjw99} Williams, R.~J.~R.\ 1999, \mnras, 310, 789 
\bibitem[Williams(2002)]{rjw02} Williams, R.~J.~R.\ 2002, \mnras, 331, 693 
\bibitem[Wise \& Abel(2008)]{wa08} Wise, J.~H., \& Abel, T.\ 2008, \apj, 684, 1 
\bibitem[Wise \& Cen(2009)]{wc09} Wise, J.~H., \& Cen, R.\ 2009, \apj, 693, 984 

\end{thebibliography}

\end{document}